\documentclass[aps,prb,twocolumn,a4paper,reprint,noeprint,xcolor=dvipsnames,superscriptaddress]{revtex4-2}
\usepackage[utf8x]{inputenc}
\usepackage{url}
\usepackage{amsmath}
\usepackage{bbold} 
\usepackage{graphicx}
\usepackage{dcolumn}
\usepackage{bm}
\usepackage{xcolor}
\usepackage{hyperref}
\usepackage{dsfont}
\usepackage[percent]{overpic}
\usepackage{empheq}
\usepackage{tikz}
\usepackage{easyReview}
\usetikzlibrary{positioning,shapes.misc,calc,3d,arrows, decorations.pathmorphing}
\usepackage{outlines}
\usepackage{subfigure}

\renewcommand{\vec}[1]{\mathbf{#1}}

\renewcommand\vec[1]{\ensuremath\boldsymbol{#1}} 

\begin{document}
\title{Competing magnetic orders and multipolar Weyl fermions in 227 pyrochlore iridates}

\author{Konstantinos Ladovrechis}
\affiliation{Institute for Theoretical Physics and W\"urzburg-Dresden Cluster of Excellence ct.qmat,\\ Technische Universit\"at Dresden, 01069 Dresden, Germany}

\author{Tobias Meng}
\affiliation{Institute for Theoretical Physics and W\"urzburg-Dresden Cluster of Excellence ct.qmat,\\ Technische Universit\"at Dresden, 01069 Dresden, Germany}

\author{Bitan Roy}
\affiliation{Max-Planck-Institut f\"ur Physik Komplexer Systeme, N\"othnitzer Str. 38, 01187 Dresden, Germany}
\affiliation{Department of Physics, Lehigh University, Bethlehem, Pennsylvania 18015, USA}

\date{\today}

\begin{abstract}
Owing to comparably strong spin-orbit coupling and Hubbard interaction, iridium based 227 pyrochlore oxides harbor a rich confluence of competing magnetic orders and emergent multipolar Weyl quasiparticles. Here we show that this family of materials is predominantly susceptible toward the nucleation of electronic noncoplanar all-in all-out (AIAO) and three-in one-out (3I1O) orders, respectively transforming under the singlet $A_{2u}$ and triplet $T_{1u}$ representations, supporting octupolar and dipolar Weyl fermions, and favored by strong on-site Hubbard and nearest-neighbor ferromagnetic interaction. Furthermore, a coplanar magnetic order generically appears as an intermediate phase between them. This order transforms under the triplet $T_{2u}$ representation and also hosts octupolar Weyl fermions. With the AIAO and 3I1O phases possibly being realized in (Nd$_{1-x}$Pr$_{x}$)$_2$Ir$_2$O$_7$ when $x=0$ and 1, respectively, the intervening $T_{2u}$ order can in principle be found at an intermediate doping ($0<x<1$) or in pressured (hydrostatic) Nd$_2$Ir$_2$O$_7$.   
\end{abstract}

\maketitle

\begin{figure}[t!]
  \includegraphics[width=0.95\linewidth]{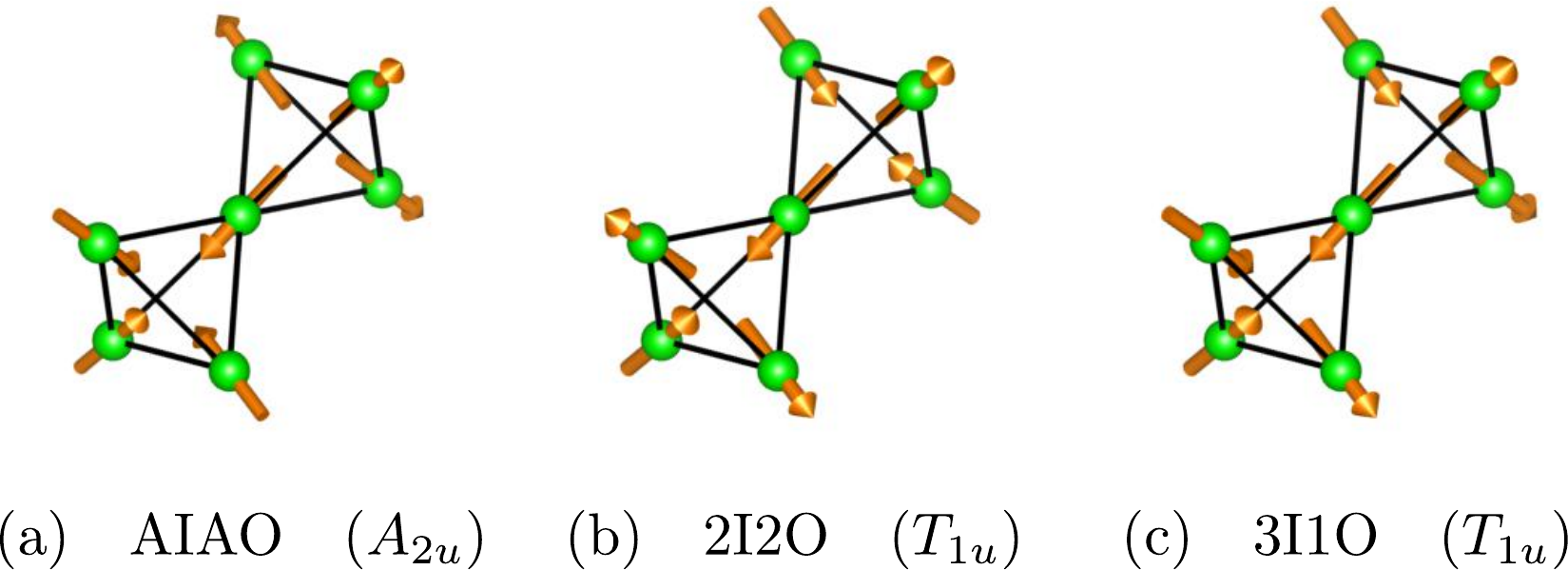}
	\includegraphics[width=0.95\linewidth]{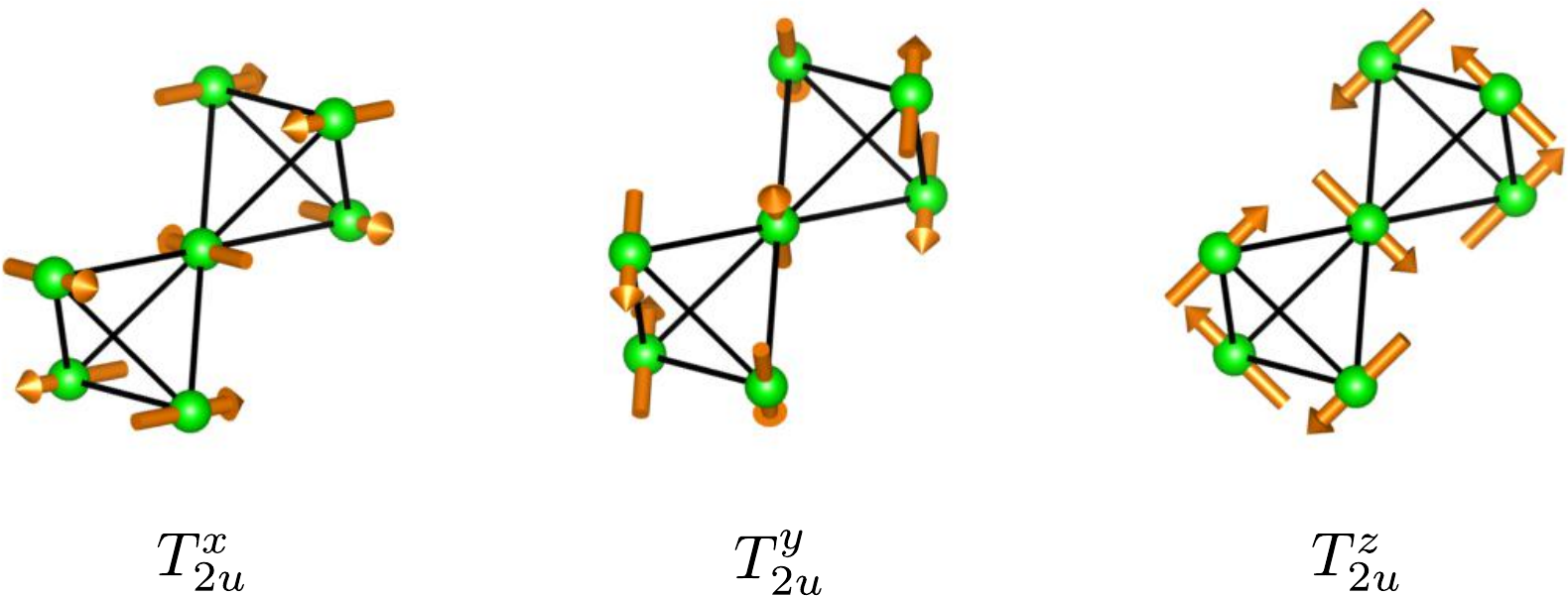}
	\caption{Noncoplanar (a) AIAO, (b) 2I2O, and (c) 3I1O arrangements of electronic spins on Ir tetrahedron. The AIAO and 2I2O orders transform under the singlet $A_{2u}$ and triplet $T_{1u}$ representation of cubic ($O_h$) point group, respectively; their coupling with the low-energy Luttinger fermions are shown in Eqs.~(\ref{eq:AIAOmatrix}) and (\ref{eq:3I1Omatrix}). The 3I1O order also transforms under $T_{1u}$ representation. Bottom: Three possible coplanar arrangements of electronic spins following the irreducible $T_{2u}$ representation. Their couplings with Luttinger fermions are shown in Eq.~(\ref{eq:T2umatrix}), with $T^{x}_{2u} \equiv T^{(1)}_{2u}$, $T^{y}_{2u} \equiv T^{(2)}_{2u}$ and $T^{z}_{2u} \equiv T^{(3)}_{2u}$.
	}~\label{fig:t2u_spins}
\end{figure}

\begin{figure}[t!]
\includegraphics[width=0.975\linewidth]{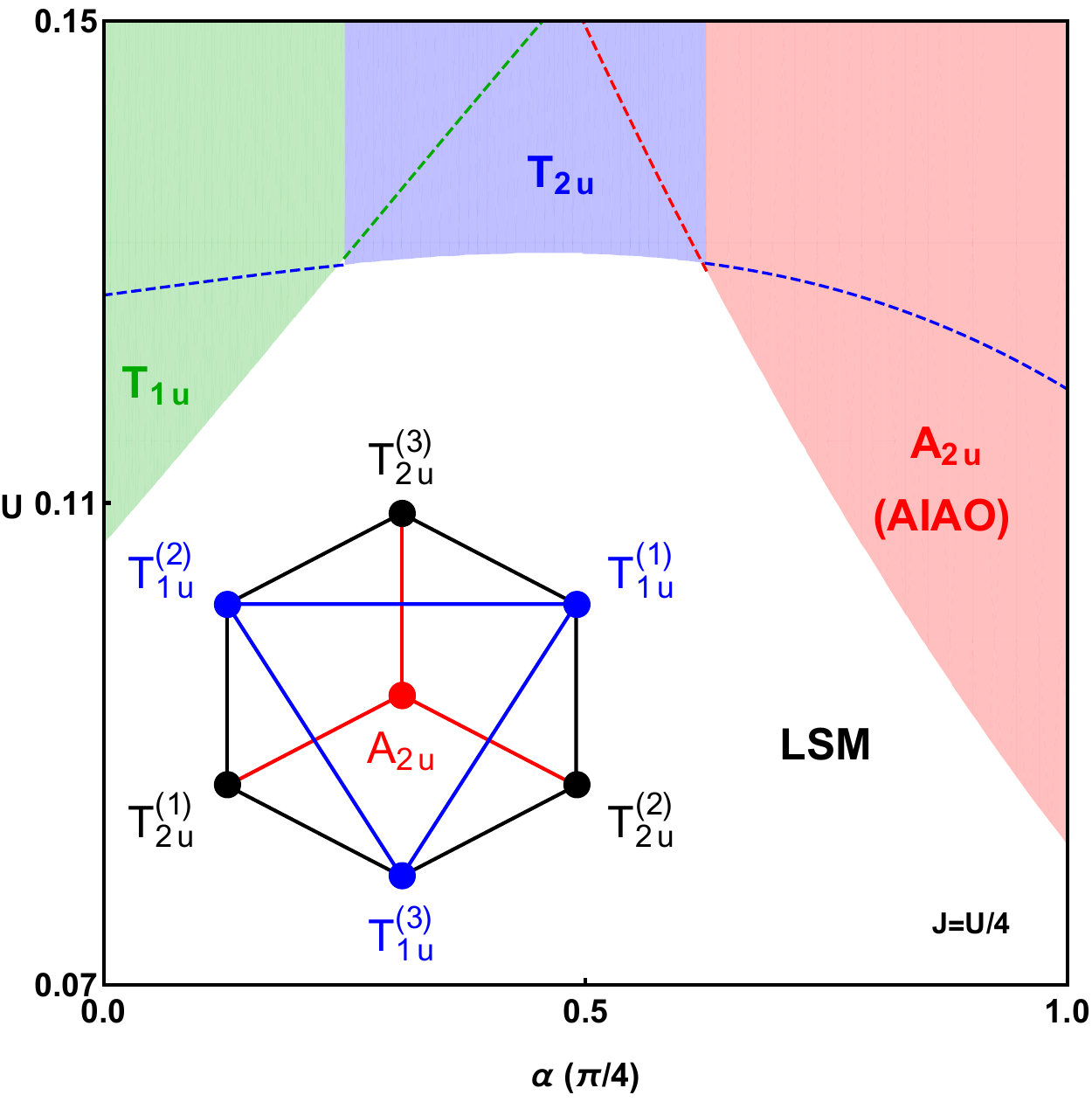}
\caption{Mean field phase diagram of an interacting Luttinger semimetal (LSM) in the presence of on-site Hubbard ($U$) and nearest-neighbor ferromagnetic ($J$) interactions for a fixed $J/U=1/4$ [Eqs.~(\ref{eq:magnetmodel})-(\ref{eq:modelcorrespondence})] at zero temperature, which should also be qualitatively applicable at sufficiently low temperatures close to the MIT. The noncoplanar $A_{2u}$ and $T_{1u}$ phases are separated by an intervening coplanar $T_{2u}$ magnetic order [Fig.~\ref{fig:t2u_spins}]. Interaction couplings are dimensionless, obtained by taking $X \Lambda(2 m_p)^{3/2}/(32 \pi^3) \to X$ for $X=U$ and $J$, where $\Lambda$ is the ultraviolet momentum cutoff, $m_p=m_1 m_2/\sqrt{m^2_1+m^2_2}$ and $\alpha=\tan^{-1}(m_1/m_2)$ [Eq.~(\ref{eq:luttinger})]. A possible coexistence between adjacent phases can be realized above the dashed lines, which we estimate from mean field susceptibilities, that, however, can be renormalized due the presence of the primary dominant order~\cite{supplementary}. Experimentally such a phase diagram can be constructed by changing the ionic radius of the Ln element~\cite{takagi2011} or applying hydrostatic or chemical pressure~\cite{tokura2015, tokura2020arXiv}. Inset: Internal algebra among the components of three magnetic orders. Six vertices (center) of the hexagon are (is) occupied by the components of $T_{1u}$ and $T_{2u}$ ($A_{2u}$) magnetic orders [Eqs.~(\ref{eq:AIAOmatrix})-(\ref{eq:T2umatrix})]. Mutually anticommuting matrix operators associated with them are connected by solid lines. When two order parameters mutually anticommute (even partially), coexistence between them gets energetically favored.             
	}~\label{fig:phasediagram}
\end{figure}

\emph{Introduction}. Experimentally accessible materials harboring emergent topological phases tend to combine two key ingredients: strong spin-orbit coupling and electronic correlation~\cite{Pesin2010, krempa-review, Jeffrau-review}. One class of materials bordering
the territory dominated by either of them are the iridium-based pyrochlore oxides Ln$_2$Ir$_2$O$_7$, also known as 227 pyrochlore iridates. Here Ln is a lanthanide element. Due to a delicate balance between the strong spin-orbit coupling and on site Coulomb repulsion among $5d$ electrons of Ir$^{4+}$ ions, 227 pyrochlore iridates exhibit rich phase diagrams featuring metal-insulator transition (MIT), competing magnetic orders hosting Weyl quasiparticles, as well as spin liquid behavior~\cite{Pesin2010, krempa-review, Jeffrau-review, vishwanath2011, takagi2011, kurita2011, krempa2012, tokura2012, Fiete2012, krempa2013, yang-nagaosa2014, savary2014, yamaji-imada2014, troyer2015, tokuraPRL2015, tokura2015, tokura2017, tokura2020arXiv, gangchen2016, Goswami2017, xidai2017, gangchen2018, bjyang2018, singh2020}. Various cuts of the global phase diagram of Ln$_2$Ir$_2$O$_7$ can be unveiled by tuning the ionic radius of the lanthanide element~\cite{takagi2011} and/or applying chemical and hydrostatic pressures~\cite{tokura2015, tokura2020arXiv}. In this Letter, we study the confluence of dominant magnetic orders in this class of materials, which involves both noncoplanar and coplanar arrangements of itinerant electronic spin of Ir$^{4+}$ ions [Fig.~\ref{fig:t2u_spins}], and construct a representative cut of the global phase diagram in the presence of both on-site Hubbard ($U$) and nearest-neighbor ferromagnetic ($J$) interactions [Fig.~\ref{fig:phasediagram}].

Most of the 227 iridates display insulating antiferromagnetic state at the lowest temperature, which possibly results from an all-in all-out (AIAO) arrangement of electronic spin on corner-shared Ir tetrahedrons~\cite{takagi2011}. One member of the familiy, Pr$_2$Ir$_2$O$_7$, in contrast remains metallic down to the lowest temperature and supports a large anomalous Hall conductivity (AHC) in the $\langle 111 \rangle$ direction, despite possessing an immeasurably small magnetic moment~\cite{pr2ir2o7:1, pr2ir2o7:2, pr2ir2o7:3}. These seemingly contradicting observations can be reconciled by noting that Pr$_2$Ir$_2$O$_7$ possibly resides at the brink of an electronic two-in two-out (2I2O) or spin-ice ordering, which ultimately produces an itinerant three-in one-out (3I1O) order that supports only a single pair of Weyl nodes in the $\langle 111 \rangle$ direction~\cite{Goswami2017}. The emergent dipolar Weyl quasiparticles then yield a large AHC without an appreciable magnetic moment. This example underlines that a systematic analysis of competing magnetic orders and emergent magnetic Weyl fermions is a worthwhile task in pyrochlore iridates.

\begin{figure}[t!]
\subfigure[]{\includegraphics[width=0.49\linewidth]{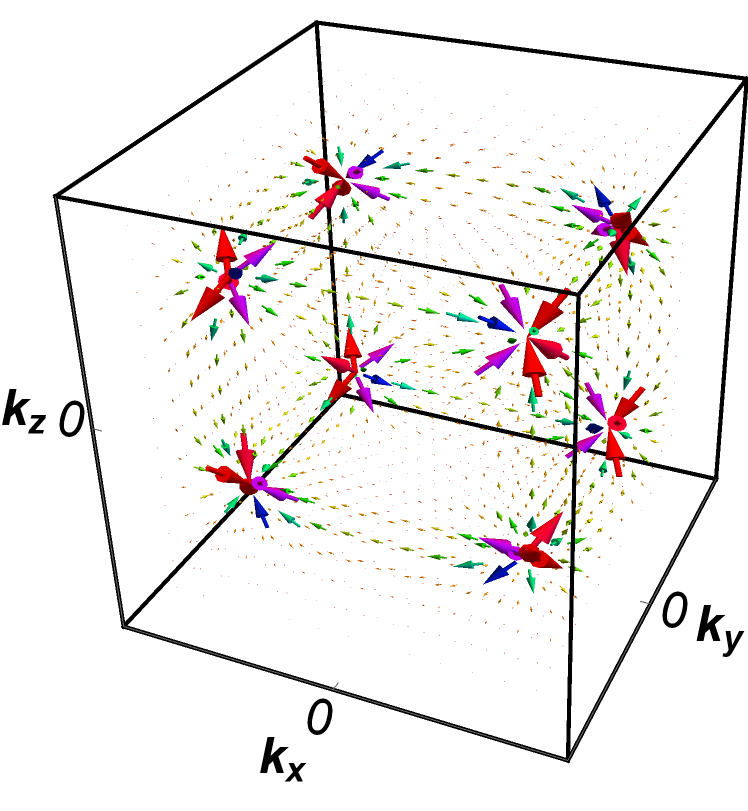}~\label{subfig:AIAO}}%
\subfigure[]{\includegraphics[width=0.49\linewidth]{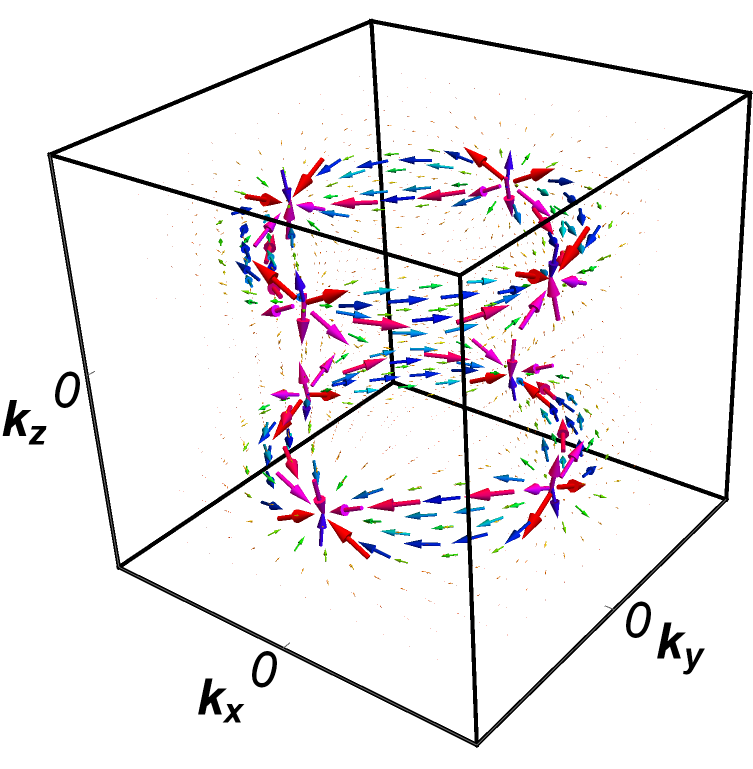}~\label{subfig:T2u}}
	\caption{Distribution of the Abelian Berry curvature for octupolar (a) $A_{2u}$ and (b) $T^z_{2u}$ magnetic orders. Weyl nodes acting as source (sink) of the Berry curvature are represented by outward (inward) arrows of the corresponding Berry flux. For both magnetic orders, eight Weyl nodes are distributed in octupolar fashions. Four of them act as source and the remaining ones as sink of the Berry curvature. The Weyl nodes and the distributions of the Berry flux in (a) and (b) are related to each other by a rotation about the $k_z$ axis by $\pi/4$. Similarly, rotation by $\pi/4$ about the $k_x$ ($k_y$) axis causes rotation between the $A_{2u}$ and $T^{x}_{2u}$ ($T^{y}_{2u}$) orders. 
	}~\label{fig:Weylnodes}
\end{figure}

In this Letter, we show that 227 pyrchlore iridates constitute a conducive platform for three competing magnetic ground states. Two of them are antiferromagnetic, transforming under the irreducible singlet $A_{2u}$ and triplet $T_{2u}$ representations, otherwise resulting from noncoplanar and coplanar arrangements of electronic spin on Ir tetrahedron, respectively, and a ferromagnetic spin-ice or 2I2O order, which ultimately gives rise to a 3I1O order. These two classes of ordered states respectively accommodate octupolar and dipolar Weyl fermions. More intriguingly, we find that the transition between the AIAO and 3I1O orders is generically mediated by an intervening $T_{2u}$ magnetic order, which, however, is shown to be \emph{topologically equivalent} to the $A_{2u}$ state [Fig.~\ref{fig:Weylnodes}]. A cut of the global phase diagram captures the competition among these ordered states [Fig.~\ref{fig:phasediagram}], which is in accordance with a proposed \emph{selection rule} among them [Fig.~\ref{fig:phasediagram}(inset)].

\emph{Model}. In Ln$_2$Ir$_2$O$_7$, one iridium ($\text{Ir}^{4+}$) atom resides at each vertex of the corner-shared tetrahedral unit cell of the pyrochlore lattice. An effective model with a single Kramers doublet at each Ir site then leads to a total of eight bands that split in a 2-4-2 pattern~\cite{kurita2011, krempa2013, yamaji-imada2014, Goswami2017}. Therefore, when the system is near half-filling, one can neglect the split-off bands and focus on the four bands close to the Fermi energy. Within this manifold, the low-energy Hamiltonian is described by so-called the Luttinger model for \emph{effective} spin-3/2 fermions~\cite{Luttinger, nagaosa-murakami-zhang, supplementary} 
\begin{equation}~\label{eq:luttinger}
	\hat{H}(\vec{k}) = \hbar^2 k^2\left[ \frac{1}{2 m_0}
	-\sum_{i=1}^{3} \frac{\hat{d}_i(\hat{\vec{k}})}{2m_1} \Gamma_i
	-\sum_{i=4}^{5} \frac{\hat{d}_i(\hat{\vec{k}})}{2m_2} \Gamma_i \right], 
 \end{equation}
where $m_0$, $m_1$ and $m_2$ bear the dimension of mass, and $\hat{\vec{d}}(\hat{\vec{k}})$ is a five-dimensional unit vector transforming in the $l=2$ representation under orbital SO(3) rotations. Its components are constructed from the spherical harmonics $Y^m_{l=2}(\theta,\phi)$. The four-component spinor basis is $\Psi^\top_{\vec{k}}= \big( c_{\vec{k},+\frac{3}{2}},c_{\vec{k},+\frac{1}{2}},c_{\vec{k},-\frac{1}{2}},c_{\vec{k},-\frac{3}{2}} \big)$, where $c_{\vec{k},m_s}$ is the fermion annihilation operator with momentum $\vec{k}$ and spin projection $m_s=\pm 3/2, \pm 1/2$. The mutually anticommuting four-component $\Gamma$ matrices are 
\begin{equation}
\Gamma_1=\kappa_3 \sigma_2, \Gamma_2=\kappa_3 \sigma_1, \Gamma_3=\kappa_2, \Gamma_4=\kappa_1, \Gamma_5=\kappa_3 \sigma_3,
\end{equation}
where Pauli matrices $\{ \kappa_\mu \}$ and $\{ \sigma_\mu \}$ with $\mu=0,\cdots, 3$ operate on the sign and magnitude of $m_s$, respectively. The Luttinger Hamiltonian emerges as an effective description near half-filling, obtained by projecting a tight-binding model of spin-1/2 electrons on pyrochlore lattice in the presence of strong spin-orbit coupling~\cite{krempa2013, Goswami2017}.

The Luttinger Hamiltonian describes a biquadratic touching of the Kramers degenerate valence and conduction bands at the $\Gamma=(0,0,0)$ point of the Brillouin zone protected by the cubic symmetry, as recently observed in Pr$_2$Ir$_2$O$_7$~\cite{pr2ir2o7:ARPES} and Nd$_2$Ir$_2$O$_7$~\cite{nd2ir2o7:ARPES}. The Kramers degeneracy of the bands is maintained by the time reversal (${\mathcal T}$) and inversion (${\mathcal P}$) symmetries. In particular,  $\vec{k} \to -\vec{k}$ and $\Psi_{\vec{k}} \to \Gamma_{13} \Psi_{-\vec{k}}$ under ${\mathcal T}$, yielding ${\mathcal T}=\Gamma_{13} {\mathcal K}$, where $\Gamma_{jk}=[\Gamma_j,\Gamma_k]/(2i)$, ${\mathcal K}$ is the complex conjugation and ${\mathcal T}^2=-1$. By contrast, $\Psi_{\vec{k}} \to \Psi_{-\vec{k}}$ under ${\mathcal P}$. In a cubic environment, $m_1 \neq m_2$ in general, where $m_1(m_2)$ is the mass parameter along the $C_{3v}$ ($C_{4v}$) axis.


\emph{Magnetic Weyl fermions}. We now discuss the prominent \emph{itinerant} magnetic orderings on the pyrochlore lattice, the corresponding reconstructed band structure, and its emergent nodal topology. In what follows we consider noncoplanar and coplanar magnetic orders of spin-1/2 electrons on pyrochlore lattice [Fig.~\ref{fig:t2u_spins}], and subsequently project them onto the Luttinger bands~\cite{supplementary}. As all magnetic orders break ${\mathcal T}$ symmetry, their coupling to the Luttinger fermions is captured by linear combinations of products of \emph{odd} numbers of spin-3/2 matrices (${\bf J}$). Onset of any magnetic order lifts the Kramers degeneracy of the bands and yields emergent Weyl nodes, at least when its amplitude is sufficiently small~\cite{comment-Weyl-insulator}.

We begin by detailing the frequently encountered AIAO arrangement of electronic spins between corner-shared Ir tetrahedra. The coupling between the AIAO magnetic order of amplitude $\xi$ with Luttinger fermions reads $ \xi \big( \Psi^\dagger \hat{A}_{2u} \Psi \big)$~\cite{vishwanath2011, savary2014, yamaji-imada2014}, where
\begin{equation}~\label{eq:AIAOmatrix}
\hat{A}_{2u} \equiv \Gamma_{45}= -\frac{2}{\sqrt{3}} \left( J_1 J_2 J_3 + J_3 J_2 J_1 \right).
\end{equation}
The AIAO order transforms under the singlet $A_{2u}$ representation of the cubic ($O_h$) point group and supports eight Weyl nodes at $(\pm 1, \pm 1, \pm 1)k^{A_{2u}}_\star$, where $k^{A_{2u}}_\star=\left[ 2 m_1 \xi/(3 \hbar^2) \right]^{1/2}$. Four of them act as sources and sinks of Abelian Berry curvature, and they are arranged in an \emph{octupolar} fashion in momentum space [Fig.~\ref{subfig:AIAO}]. The AIAO phase preserves the cubic symmetry, and the net Berry flux through any high-symmetry plane is exactly zero, yielding zero AHC, as in most of the 227 iridates.

A somewhat uncommon magnetic order results from 2I2O or spin-ice configurations of electronic spins~\cite{Goswami2017}. Their coupling with Luttinger fermions reads $\rho_j \Psi^\dagger \hat{T}^{(j)}_{1u} \Psi$, where $\rho_j$ are the amplitudes of the 2I2O orders. For $j=1,2$, and $3$, the magnetic moment points in the $\pm \hat{x}$, $\pm \hat{y}$ and $\pm \hat{z}$ direction, respectively, as
\begin{equation}~\label{eq:3I1Omatrix}
\hat{T}^{(j)}_{1u} \equiv \Gamma_{j} \Gamma_{45}= \frac{1}{3} \left( 7 J_j -4 J^3_j \right) 
\end{equation}   
and the itinerant 2I2O orders possess dominant dipole moments along the principle axes. However, in a cubic environment the magnetic moment of the itinerant spin-ice order gets locked along one of the body-diagonal $\langle 111 \rangle$ directions, causing nucleation of a triplet spin-ice or 3I1O order. Both 2I2O and 3I1O orders transform under the triplet $T_{1u}$ representation. The coupling of a 3I1O order with Luttinger fermion reads $(\rho/\sqrt{3}) \Psi^{\dagger} \left( \Gamma_1 + \Gamma_2 + \Gamma_3 \right) \Gamma_{45} \Psi$, when $\rho_1=\rho_2=\rho_3=\rho$. The 3I1O phase supports only a single pair of Weyl nodes in the body diagonal direction that are located at $\pm(1,1,1) k^{T_{1u}}_\star$, where $k^{T_{1u}}_\star= \left[ 2 m_1 \rho/(3 \hbar^2) \right]^{1/2}$. The 3I1O order consequently supports AHC in the $\langle 111 \rangle$ direction, given by $\sigma_{\langle 111 \rangle}=e^2 \sqrt{2 m_1 \rho}/(\pi \hbar^2 \sqrt{3})$~\cite{Goswami2017}, which makes it a prominent candidate for Pr$_2$Ir$_2$O$_7$~\cite{pr2ir2o7:1, pr2ir2o7:2, pr2ir2o7:3}.

Finally, we turn to the triplet magnetic order transforming under the $T_{2u}$ representation. Its coupling with Luttinger fermions reads $\phi_j \Psi^\dagger \hat{T}^{(j)}_{2u} \Psi$, where $\phi_j$ are the amplitudes of three $T_{2u}$ orders with $j=1,2,3$~\cite{supplementary}, and   
\allowdisplaybreaks[4]
\begin{align}~\label{eq:T2umatrix}
\hat{T}^{(1)}_{2u} &= \frac{J_1(J^2_2-J^2_3)}{\sqrt{3}}=\left( -\frac{1}{2}\Gamma_{15}+\frac{\sqrt{3}}{2}\Gamma_{14} \right),
\nonumber\\
\hat{T}^{(2)}_{2u} &= \frac{J_2(J^2_3-J^2_1)}{\sqrt{3}}=\left( -\frac{1}{2}\Gamma_{25}-\frac{\sqrt{3}}{2}\Gamma_{24} \right), 
\nonumber\\
\hat{T}^{(3)}_{2u} &= \frac{J_3(J^2_1-J^2_2)}{\sqrt{3}}=\Gamma_{35}.
\end{align}
The corresponding \emph{coplanar} arrangements of electronic spin on corner-shared Ir tetrahedron are shown in Fig.~\ref{fig:t2u_spins}. The emergent quasiparticle spectra inside the $T^{(3)}_{2u}$ phase are $E_{n,s}(\vec{k})=E_0(\vec{k})+ \text{sgn}(n)\epsilon_s(\vec{k})$, where $E_0(\vec{k})=\hbar^2 k^2/(2 m_0)$ captures the particle-hole anisotropy and
\allowdisplaybreaks[4]
\begin{align}
\epsilon_s(\vec{k}) &=\frac{\hbar^2}{2m_p}\bigg[C^2_\alpha d^2_3(\vec{k})+ S^2_\alpha d^2_5(\vec{k}) + \bigg\{ \text{sgn}(s)\frac{2m_p}{\hbar^2}\sqrt{3}\phi_3
\nonumber\\
&+ \bigg[ C^2_\alpha \sum_{i=1,2} d^2_i(\vec{k}) + S^2_\alpha d^2_4(\vec{k}) \bigg]^{1/2} \bigg\}^2 \bigg]^{1/2}\;,
\end{align}
with $C_\alpha=\cos \alpha$, $S_\alpha =\sin \alpha$, and $n,s=\pm$. The conduction and valence bands correspond to $n=\pm$, respectively. While the $s=+$ bands are gapped, the $s=-$ bands display touchings of Kramers nondegenerate bands at \emph{eight} isolated Weyl points in the Brillouin zone, located at $(\pm \sqrt{2},0,\pm 1) k^{T_{2u}}_\star$ and $(0,\pm \sqrt{2},\pm 1) k^{T_{2u}}_\star$, where 
\begin{equation}~\label{eq:separationWeylT2u}
k^{T_{2u}}_\star= \left[ 4m_1^2m_2^2/[3(m_1^2+2m_2^2)] \right]^{1/4}\; \big( \sqrt{\phi_3}/\hbar \big).
\end{equation}   
Four Weyl nodes act as sources and four as sinks of Berry curvature [Fig.~\ref{subfig:T2u}]. The overall octupolar arrangement of eight Weyl nodes conforms with the fact that this phase possesses only a finite octupole moment.

Despite its distinct Weyl node arrangement, we find the $T^{(3)}_{2u}$ phase to be topologically \emph{equivalent} to AIAO.  To show this, we rotate the momentum axes by $\pi/4$ about $k_z$ that takes $\vec{k} \to \vec{q}$, such that $q_x=(k_x+k_y)/\sqrt{2}$, $q_y=(k_x-k_y)/\sqrt{2}$ and $q_z=k_z$. In the new coordinate basis ($\vec{q}$), the eight Weyl nodes in the AIAO phase are located at $(\pm \sqrt{2},0,\pm 1) k^{A_{2u}}_\star$ and $(0,\pm \sqrt{2},\pm 1) k^{A_{2u}}_\star$, while those associated with the $T^{(3)}_{2u}$ order are placed at $(\pm 1,\pm 1, \pm 1) k^{T_{2u}}_\star$. The exchange of the locations of the Weyl nodes for these two orders can be further substantiated from the transformation of their respective octupole moments under the rotation by $\pi/4$ about the $z$ direction, leading to $ xyz \leftrightarrow (x^2-y^2)z$. Consequently, the octupole moment of $A_{2u}$ order (namely, $xyz$) transforms into that for the $T^{(3)}_{2u}$ order (namely, $(x^2-y^2)z$) and vice versa. Therefore, $A_{2u}$ and $T^{(3)}_{2u}$ orders are \emph{topologically equivalent}. One can show that eight Weyl nodes for the (a) $T^{(1)}_{2u}$ and (b) $T^{(2)}_{2u}$ orders are located at (a) $(\pm 1,\pm \sqrt{2},0) k^{T_{2u}}_\star$ and $(\pm 1,0,\pm \sqrt{2}) k^{T_{2u}}_\star$, and (b) $( \pm \sqrt{2}, \pm 1, 0) k^{T_{2u}}_\star$ and $(0, \pm 1,\pm \sqrt{2}) k^{T_{2u}}_\star$. These two $T_{2u}$ orders are also topologically equivalent to the $A_{2u}$ order, which can be shown by rotating the momentum axes about the $k_x$ and $k_y$ by $\pi/4$, respectively. Thus like AIAO order, $T_{2u}$ phases do not support any AHC~\cite{suzuki2019}.

We furthermore find that each microscopic spin pattern corresponding to a magnetic order in the Luttinger model is part of a family of spin patterns parametrized by $M(\alpha,\beta)=\text{diag.}[\alpha \sigma^+_{xy}+\beta\sigma^+_{yz},-\alpha \sigma^+_{xy}-\beta \sigma^-_{yz},\alpha\sigma^-_{xy}-\beta\sigma^+_{yz},-\alpha \sigma^-_{xy}+\beta \sigma^-_{yz}]$ that projects to zero when going from the eight-band model to the Luttinger Hamiltonian, where $\sigma^{\pm}_{jk}=\sigma_j \pm \sigma_k$. The resulting low-energy equivalence of microscopic magnetic patterns ties an additional connection between the $A_{2u}$ and $T_{2u}^ {(3)}$ orders: the family of magnetic patterns including the AIAO order also contains a rotated and rescaled version of the coplanar magnetic patterns that project to the $T_{2u}$-orders~\cite{supplementary}.

\emph{Competing orders}. The competition among the above magnetic orders can be captured by a minimal Hamiltonian composed of only \emph{local} (momentum-independent) four-fermion interactions
\begin{eqnarray}~\label{eq:magnetmodel}
H_{\rm int} &=& g_{_1} \left( \Psi^\dagger \hat{A}_{2u} \Psi \right)^2
+ g_{_2} \sum^3_{j=1} \left( \Psi^\dagger \hat{T}^{(j)}_{1u} \Psi \right)^2 \nonumber \\
&+& g^{(3)}_{_3} \left( \Psi^\dagger \hat{T}^{(3)}_{2u} \Psi \right)^2.
\end{eqnarray} 
The repulsive interaction $g_{_1}$ ($g_{_2}$) favors the nucleation of AIAO (3I1O) order. As the $T_{2u}$ orders possess zero dipole moment, the cubic environment does not cause any locking among its three components, unlike in the $T_{1u}$ ordered state. Therefore, the $T_{2u}$ order displays a \emph{three-fold} degenerate ground state associated with the coplanar $T^{x}_{2u}$, $T^{y}_{2u}$ and $T^{z}_{2u}$ configurations of itinerant electronic spin [Fig.~\ref{fig:t2u_spins}]. Without loss of generality, here we focus on the interaction $g^{(3)}_{_3}$ supporting $T^{z}_{2u}$ order. The bare values of these interactions are system-dependent. To build on a realistic microscopic model, we include both on-site Hubbard ($U$) and nearest-neighbor \emph{ferromagnetic} ($J>0$) interactions among itinerant spin-1/2 electrons, as longer range interactions become prominent close to the MIT. The corresponding Hamiltonian reads 
\begin{eqnarray}~\label{eq:UJmodel}
H_{\rm mic}=3U \sum^4_{i=1} S_{i} S_{i} + 3 J \sum_{i \neq j=1}^{4} S_{i} S_{j}\;,  
\end{eqnarray} 
where $S_i \equiv S_{i,111} =c^\dagger_{i,s} \left[ {\boldsymbol \sigma}_{ss^\prime} \cdot (1,1,1)^\top \right] c_{i,s^\prime}/\sqrt{3}$ is the fermionic magnetization at site $i$, fixed along the $[111]$ direction, as electronic spins in the dominant magnetic phases (such as AIAO, 2I2O and 3I1O) align in the body diagonal directions. Here $i=1,2,3,4$ correspond to four sites of an Ir tetrahedra, summation over repeated spin indices ($s,s^\prime=\uparrow, \downarrow$) is assumed, and ${\boldsymbol \sigma}=(\sigma_x, \sigma_y,\sigma_z)^\top$ is the vector Pauli matrices. Upon projecting $H_{\rm mic}$ to the low-energy spin-3/2 Luttinger manifold, we find~\cite{supplementary}  
\begin{equation}~\label{eq:modelcorrespondence}
g_{_1}=\frac{3}{2} \; g^{(3)}_{_3}=\frac{U}{4} \left(1-\frac{J}{2U} \right), \:
g_{_2}=\frac{U}{12} \left( 1+\frac{J}{6U} \right).
\end{equation}
Near half-filling, the on-site Hubbard favors antiferromagnetic $A_{2u}$ and $T_{2u}$ orders. By contrast, the ferromagnetic interaction is conducive for $T_{1u}$ order (possessing finite ferromagnetic moment).

Next, we analyze this model within the mean field approximation. To this end, we perform Hubbard-Stratonovich decoupling of quartic interactions and subsequently integrate out gapless fermions to compute the bare mean field susceptibilities for competing magnetic phases with zero external momentum and frequency. The inverse of the susceptibility yields the requisite critical couplings for the corresponding ordered states, which is finite due to the vanishing density of states ($\rho(E)\sim \sqrt{E}$) in Luttinger system. For a given set of parameters the phase with minimal critical coupling nucleates first from Luttinger semimetal~\cite{supplementary}. We follow this prescription to construct a cut of the global phase diagram in the presence of on-site Hubbard ($U$) and nearest-neighbor ferromagnetic ($J$) interactions for a fixed $J/U$, shown in Fig.~\ref{fig:phasediagram}. The phase diagram shows that the dominant AIAO and 3I1O orders are indeed separated by, and coexisting with, an intervening $T_{2u}$ order.

The arrangement of these phases can be substantiated from the internal algebra of corresponding matrix operators in the low-energy Luttinger subspace [Fig.~\ref{fig:phasediagram} (Inset)].  First, we note that the components of the $T_{1u}$ order, each representing metallic spin-ice or 2I2O order, mutually anticommute [Eq.~(\ref{eq:3I1Omatrix})]. As a result, inside the pure $T_{1u}$ ordered phase three components of 2I2O order get locked along one of the $\langle 111 \rangle$ directions, yielding the triplet spin-ice or 3I1O order, which is energetically favored over its uniaxial counterparts~\cite{2I2O:comment}. The singlet $A_{2u}$ order anticommutes with all three components of $T_{2u}$ order. Finally, each component of $T_{2u}$ order anticommutes with two components of the $T_{1u}$ order and vice versa. Therefore, in a conducive environment one expects a coexistence between $T_{2u}$ and $A_{2u}$ or $T_{1u}$ orders to be energetically favored, in qualitative agreement with our findings in Fig.~\ref{fig:phasediagram}. However, each component of $T_{2u}$ coexists with only two anticommuting components of the $T_{1u}$ order. Hence, whenever two order parameters mutually anticommute (even partially), they are expected to reside next to each other and a coexistence between them can be energetically favored. In the SM we show that even when $g^{(3)}_{3} \equiv 0$, pure repulsive interactions in the $A_{2u}$ and $T_{1u}$ channels give rise to an effective interaction in the $T_{2u}$ channel due to the Fierz relations among them~\cite{Herbut2009, Nishi2005, Jaeckel2003, szabo-moessner-roy}, yielding a intermediate $T_{2u}$ ordered phase in the $(g_{_1}, g_{_2})$ plane, see Fig.~S1 of SM~\cite{supplementary}, in agreement with the above \emph{selection rule} among competing orders.

\emph{Summary and discussions}. In summary, we show that 227 pyrochlore iridates are susceptible to three dominant magnetic phases at low temperatures: the antiferromagnetic AIAO and $T_{2u}$, and the ferromagnetic 3I1O. Respectively, these two classes of ordered phases support octupolar and dipolar Weyl fermions, but only the latter ones produce AHC in the $\langle 111 \rangle$ directions as observed in Pr$_2$Ir$_2$O$_7$~\cite{pr2ir2o7:1, pr2ir2o7:2, pr2ir2o7:3}. While individually on-site Hubbard ($U$) and nearest-neighbor ferromagnetic interaction ($J$) support AIAO and 3I1O orders, respectively, the $T_{2u}$ antiferromagnet can emerge as an intervening phase, when the strengths of $U$ and $J$ are comparable [Fig.~\ref{fig:phasediagram}]. In future, we will address such intriguing competition from an unbiased renormalization group analysis~\cite{boettcher-herbut, szabo-moessner-roy}.

The MIT in 227 pyrochlore iridates can be suppressed around a critical doping $x=0.8$ in (Nd$_{1-x}$Pr$_{x}$)$_2$Ir$_2$O$_7$ or by applying 5 GPa hydrostatic pressure on Nd$_2$Ir$_2$O$_7$~\cite{tokura2015}. Growing evidences in favor of the AIAO and 3I1O orders in Nd$_2$Ir$_2$O$_7$~\cite{nd2ir2o7:thinfilm1} and Pr$_2$Ir$_2$O$_7$~\cite{pr2ir2o7:thinfilm1}, respectively, strongly suggest that the predicted antiferromagnetic $T_{2u}$ order can be realized in (Nd$_{1-x}$Pr$_{x}$)$_2$Ir$_2$O$_7$ around $x=0.8$ or in pressured Nd$_2$Ir$_2$O$_7$ around 5 GPa. Even though both $A_{2u}$ and $T_{2u}$ orders support octupolar Weyl fermions, they can be unambiguously identified from the distinct locations of the associated Weyl nodes, yielding Fermi arc surface states connecting them~\cite{Weylreview, slager-juricic-roy-Fermiarc}, via scanning tunneling microscope, for example. Distinct octupolar moments associated with $A_{2u}$ and $T_{2u}$ orders can also be established from torque magnetometry measurements~\cite{ong:torque}.

\emph{Acknowledgments}. K.L.~and T.M.~acknowledge financial support by the Deut\-sche For\-schungs\-ge\-mein\-schaft via the Emmy Noether Programme ME4844/1-1 (project id 327807255), the Collaborative Research Center SFB 1143 (project id 247310070), and the Cluster of Excellence on Complexity and Topology in Quantum Matter \textit{ct.qmat} (EXC 2147, project id 390858490). B.R.~was supported by the Startup grant from Lehigh University.


\begin{thebibliography}{}
\bibitem{2I2O:comment} Each component of uniaxial 2I2O or spin-ice order supports a pair of Weyl nodes along one of the principle axes and a line-node in the corresponding perpendicular plane, thus yielding dominant $\rho(E) \sim |E|$ at low energies. By contrast, triplet spin-ice or 3I1O phase supports only a pair of Weyl nodes along a body diagonal, yielding $\rho(E) \sim |E|^2$, and thereby causing power-law suppression of density of states and gain in condensation energy.
\end{thebibliography}
\end{document}